\documentclass[12pt]{iopart}
\pdfoutput=1
\usepackage{graphicx}
\usepackage{bbm}
\usepackage{subfigure}
\newcommand{\bea}{\begin{eqnarray} } 
\newcommand{\eea}{\end{eqnarray} } 
\newcommand{\be}{\begin{equation} } 
\newcommand{\ee}{\end{equation} } 
\newcommand{\ltsimeq}{\raisebox{-0.6ex}{$\,\stackrel
        {\raisebox{-.2ex}{$\textstyle <$}}{\sim}\,$}}

\begin{document}
\title{Decoherence-assisted transport in quantum networks}

\author{Adriana Marais$^1$, Ilya Sinayskiy$^1$, Alastair Kay$^{2,3}$, Francesco Petruccione$^1$ and Artur Ekert$^{2,4}$}

\address{$^1$ Quantum Research Group, School of Chemistry and Physics and National Institute for Theoretical Physics,
University of KwaZulu-Natal, Durban, 4001, South Africa}
\address{$^2$ Centre for Quantum
Technologies, National University of Singapore, 3 Science Drive 2, 117543 Singapore, Singapore}
\address{$^3$ Keble College, University of Oxford, Parks Road, OX1 3PG, United Kingdom}
\address{$^4$ Mathematical Institute, University of Oxford, 24-29 St. Giles’, OX1 3LB, United Kingdom}

\ead{adrianamarais@gmail.com}

\begin{abstract}
It is shown that energy transfer in a homogeneous fully connected quantum network is assisted by a decohering interaction with environmental spins. Analytic expressions for the transfer probabilities are obtained for the zero temperature case, and the effect is shown to persist at physiological temperatures. This model of decoherence-assisted energy transfer is applied to the Fenna-Matthews-Olson complex.
\end{abstract}


\submitto{\NJP}

\section{Introduction}

Recently, evidence of quantum coherence has been detected in biological systems at physiological temperatures, including the photosynthetic light-harvesting complexes of a species of green sulphur bacteria \cite{panit10} and two species of marine cryptophyte algae \cite{collini10}; organisms which are well-adapted to photosynthesise under low-light conditions \cite{beatty05,doust06}. These results are interesting both from the perspective of quantum information processing, where a major challenge is to maintain quantum coherence in systems that unavoidably interact with an environment, and from the perspective of quantum biology, which investigates whether some aspects of the functioning of living systems can only be explained quantum mechanically. Inspired by the surprising phenomenon of quantum coherence in warm, noisy, complex and yet remarkably efficient energy transfer systems, many models of environment-assisted quantum transport have been proposed \cite{envass}, typically within approximate spin-boson models of the system. However, we show here that interaction with a more structured environment; namely a spin bath, can also assist quantum efficiency, and moreover, our model is exactly solvable in suitable limits.

\subsection{Modelling the system}

All chlorophyll-based photosynthetic organisms contain light-harvesting complexes that act as antennae, absorbing photons and transferring the resulting excitation energy to the reaction centre, where the secondary photosynthetic process of charge separation takes place. The excitation transfer happens on a scale of picoseconds and with a quantum efficiency of over 95{\%}. \cite{green03}\\
\\ 
Light-harvesting antennae consist of a variety of photoactive pigments, most commonly chlorophylls or bacteriochlorophylls, held in well-defined orientations and configurations by a scaffold of proteins \cite{kosztin08}. These (bacterio)chlorophylls have dominant absorption bands in the (near UV) blue and (near IR) red regions, hence their green colour, with the transition termed the $Q_y$ transition corresponding to longer wavelength \cite{blankenship02}. Due to very fast internal conversion of higher energy singlet states \cite{renger99}, individual pigment molecules can be approximated as two-level systems, formed by the ground state and the lowest excited singlet state, which in the case of (bacterio)chlorophyll is the $Q_y$ state \cite{kosztin08}.\\
\\
Electronic excitation energy transfer (EET) is a result of a Coulomb interaction between molecules; the electrostatic energy of one initially excited molecule can be transferred to another, initially in the ground state. The excited state is characterised by an electron-hole pair, due to the promotion of an electron from the highest occupied molecular orbital to the lowest unoccupied molecular orbital \cite{may11}. In the case of strong intermolecular coupling, superpositions of states of pairs of molecules in the ground and excited state may be formed. This delocalised state is termed a $Frenkel\  exciton$ \cite{frenkel31}, with dynamics characterised by the absence of charge transfer between molecules. In the case of photosynthetic antennae, where inter-pigment distances can be as small as 10{\AA} \cite{blankenship02}, excitation dynamics are described in terms of such excitons \cite{amerongen00}.\\
\\
Due to the timescales of photon absorption, excitation lifetime, and reaction centre reopening time in light-harvesting complexes, it is reasonable to describe the transfer dynamics under the assumption that there is at most one excitation present \cite{fassioli09}. The excitonic Hamiltonian in the single excitation subspace is given by
\be
H_{\mathrm{ex}}=\sum\limits_{j}E_{j}|j\rangle\langle j|+\sum\limits_{i\neq j}J_{ij}|i\rangle\langle j|, 
\ee
where $|j\rangle$ denotes the presence of the excitation on the two-level site $j$. The site energies of the pigments are given by $E_{j}$, and are defined as the optical transition energies at the equilibrium position of nuclei in the electronic ground state \cite{adolphs06}, while the EET couplings are given by $J_{ij}$.


\subsection{Modelling the environment}

EET is typically described in the limit of either fast or slow intramolecular relaxation as compared to intermolecular transitions \cite{may11}. F\"{o}rster theory \cite{forster65} is a non-radiative resonance transfer theory applicable to the former case; for weakly coupled molecules in the presence of strong dissipation. In this regime, the excitation energy is transferred incoherently in a hopping manner between molecules. In the latter case, the Redfield equation \cite{redfield57} is commonly employed to solve for the dynamics of the reduced density matrix of the system. In this case, the weak system-environment interaction is treated perturbatively within a Markovian approximation, i.e.$\ $where the state of the system remains uncorrelated with the environment \cite{breuer02}.\\
\\
In the case of photosynthetic light-harvesting complexes, however, the reorganisation energies of the protein molecules appear to lie within the range of electronic coupling strengths between pigments \cite{adolphs06,cho05}. Thus, the EET dynamics are in the intermediate regime between the two limits, and should be investigated within the theory of open quantum systems in the non-Markovian regime, a topic which has recently attracted much attention \cite{cui08,cui09,breuer09,rivas10}. The quantum dynamics of excitons within a protein medium are typically treated within a spin-boson model of the system. Although theoretical developments have led to higher-order approximate solutions for systems that interact strongly with environmental vibronic modes, including in the non-Markovian regime \cite{ishizaki09}, such models are generally not exactly solvable. Spin baths, on the other hand, naturally describe a set of localised environmental modes \cite{prokof00,yang08}. Moreover, the interaction of a central spin with a spin bath often leads to strong non-Markovian behavior of the central spin \cite{breuer04}.\\
\\
Here, we do not attempt to provide a model of energy transfer in real photosynthetic systems that can explain the lifetime of the observed quantum coherence. A recent article \cite{chin12} reviews progress on this question. Rather, we show with analytical expressions that interaction with a finite spin bath can also assist quantum efficiency, that this effect persists at physiological temperatures, and therefore that a sufficiently large spin bath could provide a sufficiently realistic model for environments occurring in photosynthetic light-harvesting complexes and biological systems in general.\\
\\
We make the additional simplification of considering a pure dephasing interaction between the system and the environment, with the aim of an analytical description of the EET dynamics. Since the transfer efficiencies are so high, it seems reasonable to assume that the decoherent dynamics do not change the excitation number (although models that can tolerate a changing excitation number on the system spins do exist \cite{burgarth06}). 



\subsection{Decoherence-assisted transport}

In a recent article \cite{sinay12} analytic expressions were derived for the transition probability in a dimer system under the influence of a decoherent interaction with environmental spins. It was shown that there exist biologically relevant parameter regimes where such an interaction shifts the energy levels of the dimer, such that resonance and near-perfect transfer is achieved, and moreover that these effects persist at physiological temperatures; i.e. transfer probabilities of over 85\% can be achieved at 300K.\\
\\
This idea has recently been extended to the study of the noise-induced properties of a two-level system in a spin bath which undergoes a quantum phase transition \cite{giorgi12}. Here, we extend the model of decoherence-assisted transport to more complex networks.\\
\\
In section 2 we review the case of the dimer with its levels coupled to spin baths. In section 3 we analyse the dynamics of a single excitation in a fully-connected $N$-site network with sites interacting with environmental spins at zero temperature. We show analytically that the maximum probability of transfer through the network can be increased as a result of decoherent coupling to spin environments, and find that there are cases where transfer can be guaranteed. Furthermore, we show that these effects persist at physiological temperatures. In section 4 we consider as an example the Fenna-Matthews-Olson (FMO) complex \cite{fenna75}, the antenna complex found in the photosynthetic light-harvesting units of green sulphur bacteria. The FMO complex is modelled as a fully connected 7-site network with site energies and EET couplings calculated by Adolphs and Renger \cite{adolphs06}, and it is demonstrated numerically that EET is significantly assisted by decoherent interaction between each network site and a respective spin environment at a physiological temperature of 300K. 

\section{Decoherence-assisted transport in a dimer system}

For a dimer with Hamiltonian $H_{d}=\varepsilon_{1}|1\rangle\langle1|+\varepsilon_{2}|2\rangle\langle2|+J(|1\rangle\langle2|+|2\rangle\langle1|)$, the maximum transfer probability for a single excitation $\mbox{Max}[P_{1\rightarrow2}(t)]$ is given by $J^{2}/(J^{2}+\Delta^{2})$ where $J$ is the amplitude of transition, and the detuning $\Delta$ is given in terms of the energy levels of the dimer as $(\varepsilon_2-\varepsilon_1)/2$. Certain transfer is achieved when $\Delta=0$ at time $t=\pi/(2J)$, i.e. when there is resonance between the energy levels in the system.\\
\\
In a recent article \cite{sinay12}, it was shown that there exist well-defined ranges of parameters for which a pure dephasing interaction with environmental spins in a spin star configuration \cite{breuer04} assists energy transfer in the dimer system. For a dimer with each level coupled to a spin bath at zero temperature, the Hamiltonian of the total system is given by
\be
H= H_{d} + H_{B} + H_{I}.
\ee 
Each environment $B_j$ consists of $n_j$ spin-half particles
\be
H_{B}=\sum\limits_{j=1}^{2}H_{B_j}=\sum\limits_{j=1}^{2}\sum_{k=1}^{n_j}\alpha_{j}\frac{\sigma_z^{k,j}}{2},
\ee 
where $\sigma_z$ are Pauli matrices, the baths are labelled by $j=1,2$, and the spins in each bath by $k=1,...,n_{j}$. The purely decoherent interaction between each site $j$ in the system and the corresponding spin bath is modelled by 
\be
H_{I}=\sum\limits_{j=1}^{2}H_{I_j}=\sum\limits_{j=1}^{2}\sum_{k=1}^{n_j}\gamma_j|j\rangle\langle j|\frac{\sigma_z^{k,j}}{2}.
\ee
We consider the zero temperature case, and therefore the state of each bath is a pure state, the ground state, described in the collective operator basis by
\be
|\psi_{B_{j}}(0)\rangle = |\frac{n_j}{2},-\frac{n_j}{2}\rangle,
\ee
(see \cite{sinay12} for details). The Hamiltonian of the environment $H_B$ commutes with the Hamiltonian of interaction $H_I$ and therefore the state of the total system is always in a product state of the network and the baths. As a result, the effective Hamiltonian for the total system is given by 
\be
H=\sum\limits_{j=1}^{2}\varepsilon'_{j}|j\rangle\langle j|+\sum_{i,j=1\atop i\neq j}^{2}J|i\rangle\langle j|,
\ee
where $\varepsilon'_{j}=\varepsilon_{j}-\gamma_{j}n_{j}/2$.\\
\\
For the Hamiltonian $H$, the maximum transfer probability $\mbox{Max}[P_{1\rightarrow2}(t)]$ is given by $J^2/(J^2+\Delta'^2)$ where in this case the detuning is given by $\Delta'=(\varepsilon'_2-\varepsilon'_1)/2$. Certain transfer is similarly achieved when $\Delta'=0$, which in this case is possible for a wide range of parameters $\gamma_{j}$ and $n_{j}$.\\
\\ 
At zero temperature, the total state is always a product of the dimer state and bath state. This means that the dimer is always in a pure state. At non-zero temperature, however, the state of the dimer is described by the density matrix obtained by tracing out the degrees of freedom of the bath. The initial state of the bath is given by the canonical distribution
\be
\rho_{B}(0)=\prod\limits_{i=1}^{2}\frac{1}{Z_{i}}\mbox{e}^{-\beta H_{B_{j}}},
\label{canon}
\ee 
where $Z_{i}$ is the partition function of the corresponding bath
\bea
Z_{i}&=&\sum\limits_{j_{i}=0}^{n_{i}/2}\sum\limits_{m_{i}=-j_{i}}^{j_{i}}\nu(n_{i},j_{i})\langle j_{i},m_{i}|\mbox{e}^{-\beta\alpha_{i}\sigma_{i}^{z}}| j_{i},m_{i}\rangle\nonumber\\
&=&\sum\limits_{j_{i}=0}^{n_{i}/2}\nu(n_{i},j_{i})\frac{\sinh{\beta\alpha_{i}(j_{i}+1/2)}}{\sinh{\beta\alpha_{i}/2}},
\label{part}
\eea
where $\beta$ is the inverse temperature, and $\nu(n_{i},j_{i})$ denotes the degeneracy of the spin bath \cite{deg}. The effect of decoherence-enhanced transport is shown to persist at a physiological temperature of 300K, where transfer probabilities of nearly 90$\%$ can be achieved in the dimer for biologically relevant parameters \cite{sinay12}.

\section{Decoherence-assisted transport in a fully-connected quantum network}

\subsection{The fully symmetric network}

For a fully connected network of $N$ qubits interacting via homogeneous $XX$ coupling with coupling strength $J/2$ and with equal site energies $\varepsilon$, the effective Hamiltonian in the single excitation subspace is given by
\be
H_{N} = \sum_{i=1}^{N}\varepsilon|i\rangle\langle i| + \sum_{i,j=1\atop i\neq j}^{N}J|i\rangle\langle j|. 
\ee
The eigenvalues for the Hamiltonian $H_{N}$ are given by
\bea
\lambda_{1,...,N-1}&=&\varepsilon-J,\\
\lambda_{N}&=&\varepsilon+(N-1)J.
\eea
The properties of the constant $\omega=\exp(i2\pi/N)$, including 
\be
\sum\limits_{j=1}^{N}\omega^{j}=0,
\ee
allow the following choice of eigenbasis for the system:
\bea
|\lambda_{m=1,...,N-1}\rangle&=&\frac{1}{\sqrt{N}}\sum\limits_{j=1}^{N}\omega^{mj}|j\rangle,\\
|\lambda_{N}\rangle&=&\frac{1}{\sqrt{N}}\sum\limits_{j=1}^{N}|j\rangle.
\eea
Each site in the network $|j\rangle$ for $j=1,...,N$ can then be written in the eigenbasis as 
\be
|j\rangle=\frac{1}{\sqrt{N}}(|\lambda_N\rangle+\sum\limits_{m=1}^{N-1}\omega^{-mj}|\lambda_{m}\rangle),
\label{sites}
\ee
and the probability of transfer of the excitation from some initial site $|I\rangle$ to a final site $|F\rangle$, both of which can be written in the form of (\ref{sites}), is then given by
\bea
P_{I\rightarrow F}(t)&=&|\langle F|\mbox{e}^{-iH_{N}t}|I\rangle|^2\nonumber\\
&=&\frac{1}{N^2}|\mbox{e}^{-it\lambda_{N}}+\mbox{e}^{-it\lambda_{1}}\sum\limits_{m=1}^{N-1}\omega^{m(F-I)}|^2\nonumber\\
&=&\frac{1}{N^2}|1-\mbox{e}^{-itNJ}|^2.
\eea
In this case the maximum probability of purely coherent transfer through the network is
\be
\mbox{Max}[P_{I\rightarrow F}(t)] =\frac{4}{N^2}, 
\ee
at time $t=\pi/(NJ)$ (and subsequent periodic revival times).\\
\\ 
Our study will focus on whether a decoherent interaction between a fully-connected network and environmental spins can enhance energy transport through the network, by breaking the symmetry properties that prevent distinguishing between one target site and another.

\subsection{Adding environmental spins}

By coupling sites in the fully connected network to independent spin environments in symmetric star configurations, the Hamiltonian of the total system is given by
\be
H = H_{N} + H_{B} + H_{I},
\ee 
with $H_{B}$ and $H_{I}$ defined as previously in section 2, but with $j=1,...,N$. We can then write the effective Hamiltonian for the total system, with arbitrary coupling of the zero temperature spin baths as 
\be
H=\sum_{j=1}^{N}\varepsilon_{j}|j\rangle\langle j| + \sum_{i,j=1\atop i\neq j}^{N}J|i\rangle\langle j|,
\label{Hk}
\ee
where $\varepsilon_{j}=\varepsilon-\gamma_{j}n_{j}/2$.\\
\\
If we select that $k$ sites are not coupled to any baths (or the same number of baths), then the resultant Hamiltonian $H_{k}$ has $k-1$ degenerate eigenvalues $\lambda_{1,...,k-1}=\varepsilon-J$, while the remaining eigenvalues are given by solutions to the polynomial of order $N-k+1$ generated by the Hamiltonian. The orthonormal eigenvectors corresponding to the degenerate eigenvalues $\lambda_{1,...,k-1}$ are given by
\be
|\lambda_m\rangle=\frac{1}{\sqrt{k}}\sum\limits_{j=1}^{k}\omega^{mj}|j\rangle
\label{eigk}
\ee
for $m=1,...,k-1$, where $\omega=\exp(i2\pi/k)$. \\
\\
The symmetry of $H_{k}$ and the resulting form of the eigenvectors (\ref{eigk}) make it possible to concentrate on the essential dynamics through a partial diagonalisation $\tilde{H}_{k}=U_{k}H_{k}U_{k}^{\dagger}$. The unitary transformation $U_{k}$ is given by
\be
U_{k}=|1\rangle\langle\phi|+\sum_{n=1}^{N-k}|n+1\rangle\langle n+k|+\sum_{m=1}^{k-1}|N-k+m+1\rangle\langle\lambda_m|,
\label{U}
\ee
where $|\phi\rangle$ is defined as
\be
|\phi\rangle=\frac{1}{\sqrt{k}}\sum\limits_{j=1}^{k}|j\rangle.
\label{phi}
\ee
An equivalent transformation can be applied to each set of spins with the same energies $\varepsilon_{j}$. Using the above general formalism, we now derive expressions for the transfer probabilities for some particular cases.

\subsubsection{Adding one spin bath.}

In the case where just one spin bath is coupled to a network site, the Hamiltonian $H_{N-1}$ is given by (\ref{Hk}) with $k=N-1$. In this case there are $N-2$ degenerate eigenvalues of energy $\varepsilon-J$, while the remaining eigenvalues satisfy
\be
\lambda_{N-1}+\lambda_{N}=\varepsilon+\varepsilon_{1}+(N-2)J.
\ee
If the bath is coupled to either the initial site $|I\rangle$ or the final site $|F\rangle$, the dynamics are constrained to a two-dimensional subspace which is independent from the degenerate subspace of dimension $N-2$ in the effective Hamiltonian picture. In this case, where the energy of the site coupled to the bath is $\varepsilon_{1}$, the rotated Hamiltonian $\tilde{H}_{N-1}=U_{N-1}H_{N-1}U_{N-1}^{\dagger}$ is given by
\be
\tilde{H}_{N-1}=
\left( {\begin{array}{ccccccc}
\varepsilon+(N-2)J&\sqrt{N-1}J&0&0&...&0\\
\sqrt{N-1}J&\varepsilon_{1}&0&0&...&0\\
0&0&\varepsilon-J&0&...&0\\
0&0&0&\varepsilon-J&&0\\
\vdots&\vdots&\vdots&&\ddots\\
0&0&0&0&&\varepsilon-J\\
 \end{array} } \right).\\
\ee
As in the case of the dimer, transfer is optimised by inducing resonances; setting the energy level of the bath site to $\varepsilon_{1}=\varepsilon+(N-2)J$ realises the maximum probability of transfer
\be
\mbox{Max}[P_{I\rightarrow F}(t,\varepsilon)]=\frac{1}{N-1},
\ee
at time $t=\pi/(NJ)$ (and subsequent periodic revival times). The situation where the bath is coupled to neither site $|I\rangle$ nor $|F\rangle$ is a special case of the model analysed in section 3.2.3.

\subsubsection{Adding two spin baths.}

In the case where two network sites are coupled to spin baths, the Hamiltonian $H_{N-2}$ is given by (\ref{Hk}). In this instance, there are $N-3$ degenerate eigenvalues of energy $\varepsilon-J$, while $\lambda_{N-2}=\varepsilon_{1}-J$ and the remaining eigenvalues satisfy
\be
\lambda_{N-1}+\lambda_{N}=\varepsilon+\varepsilon_{1}+(N-2)J.
\ee
If the baths are coupled to the initial and final sites, $|I\rangle$ and $|F\rangle$, the dynamics are constrained to a three-dimensional subspace which is independent from the degenerate subspace of the effective Hamiltonian $\tilde{H}_{N-2}=U_{N-2}H_{N-2}U_{N-2}^{\dagger}$
\be
\tilde{H}_{N-2}=
\left( {\begin{array}{llllll}
\varepsilon+(N-3)J&\sqrt{N-2}J&\sqrt{N-2}J&0&...&0\\
\sqrt{N-2}J&\varepsilon_{1}&J&0&...&0\\
\sqrt{N-2}J&J&\varepsilon_{2}&0&...&0\\
0&0&0&\varepsilon-J&&0\\
\vdots&\vdots&\vdots&&\ddots\\
0&0&0&0&&\varepsilon-J\\
 \end{array} } \right).\\
\ee
The probability of transfer between sites $|I\rangle$ and $|F\rangle$ (with effective site energies $\varepsilon_{1}$ and $\varepsilon_{2}$) is given by
\be
P_{I\rightarrow F}(t)=\left|\sum\limits^{N}_{j=N-2}c_{j}\mbox{e}^{-it\lambda_{j}}\right|^2\leq(|c_{N}|+|c_{N-1}|+|c_{N-2}|)^2.
\ee
The constants $c_{j}=\langle F|\lambda_j\rangle\langle \lambda_j|I\rangle$ are given by
\bea
c_{1,...,N-3}&=&0\\
c_{N-1}+c_{N}&=&-c_{N-2}=\frac{J(J-\varepsilon_2+\lambda_{N-2})}{(\lambda_{N-2}-\lambda_{N-1})(\lambda_{N-2}-\lambda_{N})}.
\eea
Clearly, perfect transfer can only be achieved if $|c_{N-2}|=1/2$. Setting $\varepsilon_{1}=\varepsilon_{2}$ enables this, and from the resulting form of the eigenvector $|\lambda_{N-2}\rangle=1/\sqrt{2}(|I\rangle-|F\rangle)$, the existence of times for which perfect transfer is achieved follows readily.\\
\\
We now show that such an effect persists at physiological temperature by considering the Hamiltonian $H_{N-2}$ with spin baths at a temperature of 300K coupled to the initial and final sites.
\begin{figure}
   \centering
    \includegraphics[scale=0.5]{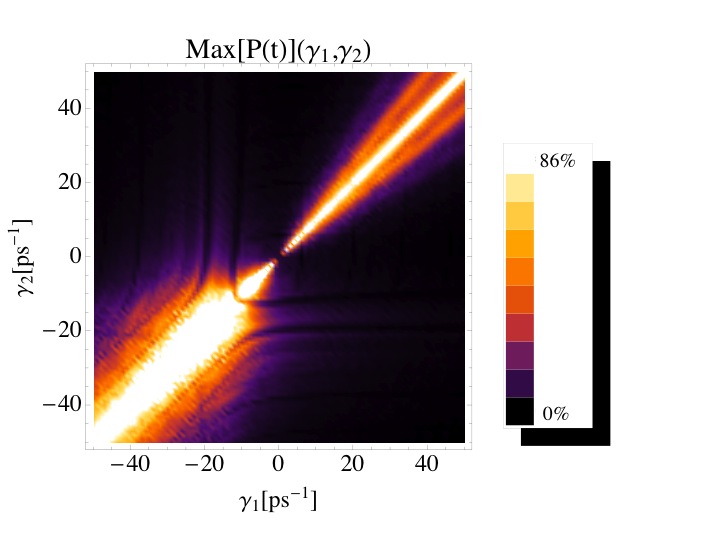}
     \caption{Graph of the maximum of the probability of transition $\mbox{Max}[P_{I\rightarrow F}(t)]$ at 300K in a 10-site fully connected homogeneous network, with equal isolated site energies, intersite coupling $J=10$ps$^{-1}$, and spin baths coupled to each of the initial and final sites, with $n_1=n_2=10$ and bath energy parameter $\alpha_{1}=\alpha_{2}=150$ps$^{-1}$.}
\end{figure}
In this case the initial state of the bath is given by the canonical distribution defined in (\ref{canon}), with corresponding partition function defined in (\ref{part}), and the density matrix describing the state of the network is obtained by tracing out the degrees of freedom of the bath.\\
\\
In figure 1 the maximum of the probability of transition $\mbox{Max}[P_{I\rightarrow F}(t)]$ for such a system is plotted as a function of the coupling constants $\gamma_1$ and $\gamma_2$ with spin baths coupled to each of the initial and final sites. It can be seen that in regions where $\varepsilon_1=\varepsilon_2\neq\varepsilon$, transfer probabilities of up to 86$\%$ are achieved, and the breadth of the region indicates the effect's robustness against imperfections.

\subsubsection{Spin baths on intermediate network sites.}

We now consider the case where $N-k$ of the fully connected network sites, sites $|k+1\rangle,...,|N\rangle$, are coupled to spin baths, but now the initial and final sites are chosen from any of the sites $|1\rangle,...,|k\rangle$. The initial and final sites $|I\rangle$ and $|F\rangle$ can then be written in the form of (\ref{sites}), where $j\in\{1,...,k\}$, as
\be
|j\rangle=\frac{1}{\sqrt{k}}(|\phi\rangle+\sum\limits_{m=1}^{k-1}\omega^{-mj}|\lambda_{m}\rangle).
\ee
Note that the state $|\phi\rangle$ defined in (\ref{phi}) is not an eigenstate of the system. The probability of transfer of the excitation from the initial to final site is then given by
\bea
P_{I\rightarrow F}(t)&=&|\langle F|\mbox{e}^{-itH_{k}}|I\rangle|^2\nonumber\\
&=&\frac{1}{k^2}|\langle\phi|\mbox{e}^{-itH_{k}}|\phi\rangle+\mbox{e}^{-it(\varepsilon-J)}\sum\limits_{m=1}^{k-1}\omega^{m(F-I)}|^2\nonumber\\
&=&\frac{1}{k^2}|1-\mbox{e}^{it(\varepsilon-J)}\langle\phi|\mbox{e}^{-itH_{k}}|\phi\rangle|^2\nonumber\\
&=&\frac{1}{k^2}|1-\mbox{e}^{it(\varepsilon-J)}\langle 1|\mbox{e}^{-it\tilde{H}_{k}}|1\rangle|^2.
\label{genpr}
\eea
The maximum possible value of the above expression for the transfer probability is 
\be
\mbox{Max}[P_{I\rightarrow F}(t)]=\frac{4}{k^{2}},
\ee
and this value is approximately achieved for almost all systems with Hamiltonian $H_{k}$ independent of the details of the $\{\varepsilon_{j}\}$ after sufficiently long times (see Appendix for further details). The motivation for this statement is that any system always exhibits perfect revivals, i.e. there always exists a time such that $\mbox{e}^{-itH_{k}}|\phi\rangle$ returns $|\phi\rangle$.\\
\\
\begin{figure}
   \centering
    \includegraphics[scale=0.5]{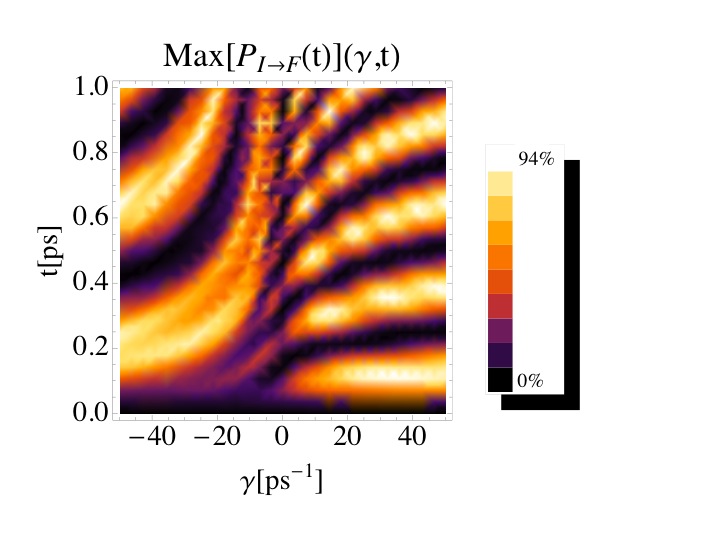}
     \caption{Graph of the probability of transition $P_{I\rightarrow F}(\gamma,t)$ at 300K in a 4-site fully connected homogeneous network, with equal isolated site energies,  intersite coupling $J=10$ps$^{-1}$, and spin baths coupled to the intermediate sites, with $n_1=2,\ n_2=8$, $\gamma_1=\gamma_2=\gamma$ and bath energy parameter $\alpha_{1}=\alpha_{2}=150$ps$^{-1}$.}
\end{figure}
This is an increase in the maximum probability of transfer over time which is given by $4/N^{2}$ in the fully symmetric case. In the case where $k=2$, when all but the initial and final sites are coupled to spin baths, it can be seen that perfect transfer is achieved.\\
\\
We now show that such an effect persists at physiological temperature by considering the Hamiltonian $H_{2}$ with spin baths at a temperature of 300K coupled to intermediate sites, i.e. all but the initial and final sites. The initial states of the baths are defined as previously.\\
\\ 
In figure 2, the probability of transition $P_{I\rightarrow F}(t)$ for such a system is plotted as a function of the coupling constant $\gamma$, with different spin baths coupled to each of the intermediate sites, and isolated site energies all equal. It can be seen that for all $\gamma\neq0$, i.e. for all effective intermediate site energies not equal to the energy of the initial and final sites, there exist times where transfer probabilities of up to 94$\%$ are achieved.

\section{Decoherence-assisted transport in the FMO complex}

The first evidence of quantum coherence in photosynthetic antennae at physiological temperature was detected in green sulphur bacteria and cryptophyte algae \cite{panit10,collini10} which are both organisms able to photosynthesise efficiently at low light intensities \cite{doust06}; a species of green sulfur bacteria has even been found living at a depth of 2500m in the Pacific Ocean near a thermal vent \cite{beatty05}. Green sulphur bacteria uniquely contain a complex called the Fenna-Matthews-Olson (FMO) complex \cite{fenna75}. The FMO pigment-protein complex mediates excitation energy transfer from the large main antenna system of green sulphur bacteria, the chlorosome, to the reaction centre \cite{blankenship04}. The structure of the FMO complex was first resolved in three-dimensions using X-ray crystallography in 1975 \cite{fenna75}, where it was shown to consist of three identical subunits each containing seven bacteriochlorophyll a (BChl a) pigments and enclosed within an envelope of protein. Due to the weakness of electronic coupling between pigments in different subunits, it is reasonable to consider the EET dynamics within one 7-site subunit \cite{adolphs06}.\\ 
\\
The site energies and optical transition energies (defined in section 1.1) for the FMO complex of \textit{Chlorobium tepidum}, a model organism of green sulphur bacteria, were calculated by Adolphs and Renger \cite{adolphs06}
\be
H^{\mathrm{FMO}}_{\mathrm{ex}}=
\left( {\begin{array}{ccccccc}
 200&-96&5&-4.4&4.7&-12.6&-6.2\\
 -96&320&33.1&6.8&4.5&7.4&-0.3\\
 5&33.1&0&-51.1&0.8&-8.4&7.6\\
 -4.4&6.8&-51.1&110&-76.6&-14.2&-67\\
 4.7&4.5&0.8&-76.6&270&78.3&-0.1\\
 -12.6&7.4&-8.4&-14.2&78.3&420&38.3\\
 -6.2&-0.3&7.6&-67&-0.1&38.3&230
 \end{array} } \right),
\ee
in units of cm$^{-1}$, where the zero of energy has been shifted by 12 210cm$^{-1}$. Energy transfer through the FMO complex is estimated to take place over a maximum of 5ps \cite{adolphs06}.\\\\

\begin{figure}
\centering
\includegraphics[scale=0.5]{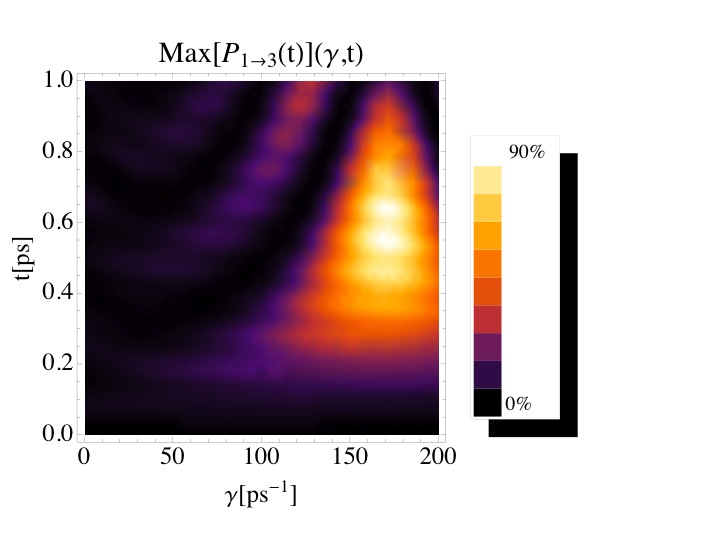}
\caption{The probability of transfer $P_{1\rightarrow 3}(\gamma,t)$ for
the FMO complex with sites coupled to a spin baths at 300K with numbers of spins at each site $n_1=2, n_4=8$ and $n_{2,3,5,6,7}=0$ and spin bath energy constant $\alpha$=150ps$^{-1}$, which is the best spin distribution for a total of 10 spins.} 
\end{figure}

The efficiency with which the FMO complex transfers excitation energy to the reaction centre plays a crucial role in the organism's survival under extremely low light conditions. However, quantum coherent EET through the bare excitonic system without adding environmental contributions to the Hamiltonian $H^{\mathrm{FMO}}_{\mathrm{ex}}$ happens with a low probability: for the transfer of the excitation to site 3 from an initial position at site 1(6) \cite{adolphs06}, the probability of transfer is just 5.0(1.3)$\%$. We now investigate the effect of a decoherent interaction with environmental spins at 300K on the process of EET in the FMO complex.\\
\\
Distributing a total of 10 environmental spins amongst the 7 sites, with the simplification that there are only even numbers of spins on each site (see \cite{deg}), gives a total of 462 different spin distributions. With the spin baths at a temperature of 300K, we calculate the maximum probability of transfer to site 3 during the first picosecond (a time we found sufficient to find the maximum), for equal environmental couplings $\gamma$ ranging between 0 and 200ps$^{-1}$ at each site. Note that here $\hbar\sim$5.3cm$^{-1}$ps.\\
\\
When the excitation is initialised at site 1, the best spin distribution is $n_1=2, n_{4}=8$ and $n_{2,3,5,6,7}=0$ (see figure 3), while for the initial site 6, the best spin distribution is $n_6=4, n_{4}=6$ and $n_{1,2,3,5,7}=0$, with the following maximum probabilities:
\bea
\mbox{Max}[P_{1\rightarrow3}]=90\%\\
\mbox{Max}[P_{6\rightarrow3}]=80\%.
\eea
For 69$\%$(60$\%$) of the spin distributions in the case of site 1(6) as the initial site, there is an increase in the transfer probability from the case with no spin baths, where the probability is 5.0$\%$(1.3$\%$) (see figure 4). Therefore, in the vast majority of cases, decoherence assists the efficiency of quantum coherent EET in the FMO complex.\\ 
\\
\begin{figure}
\centering
\mbox{\subfigure{\includegraphics[scale=0.51]{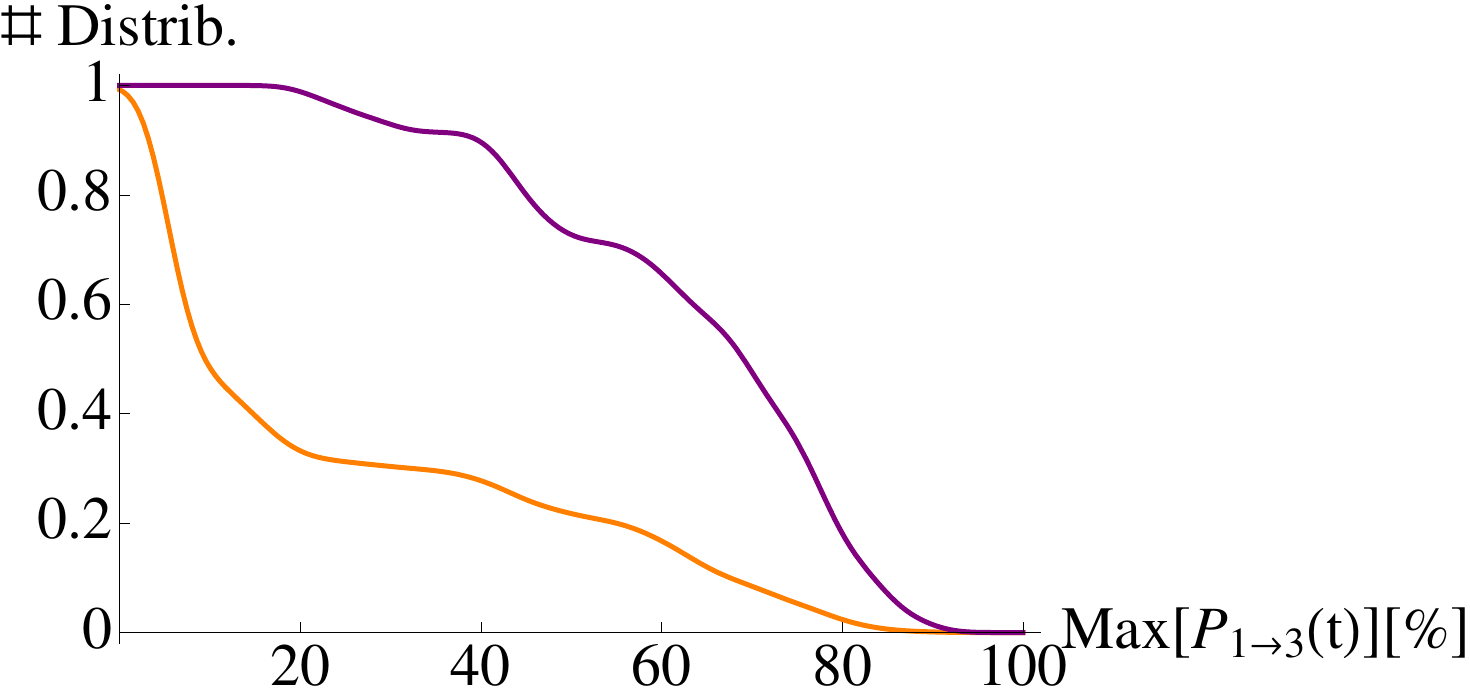}}\quad
\subfigure{\includegraphics[scale=0.51]{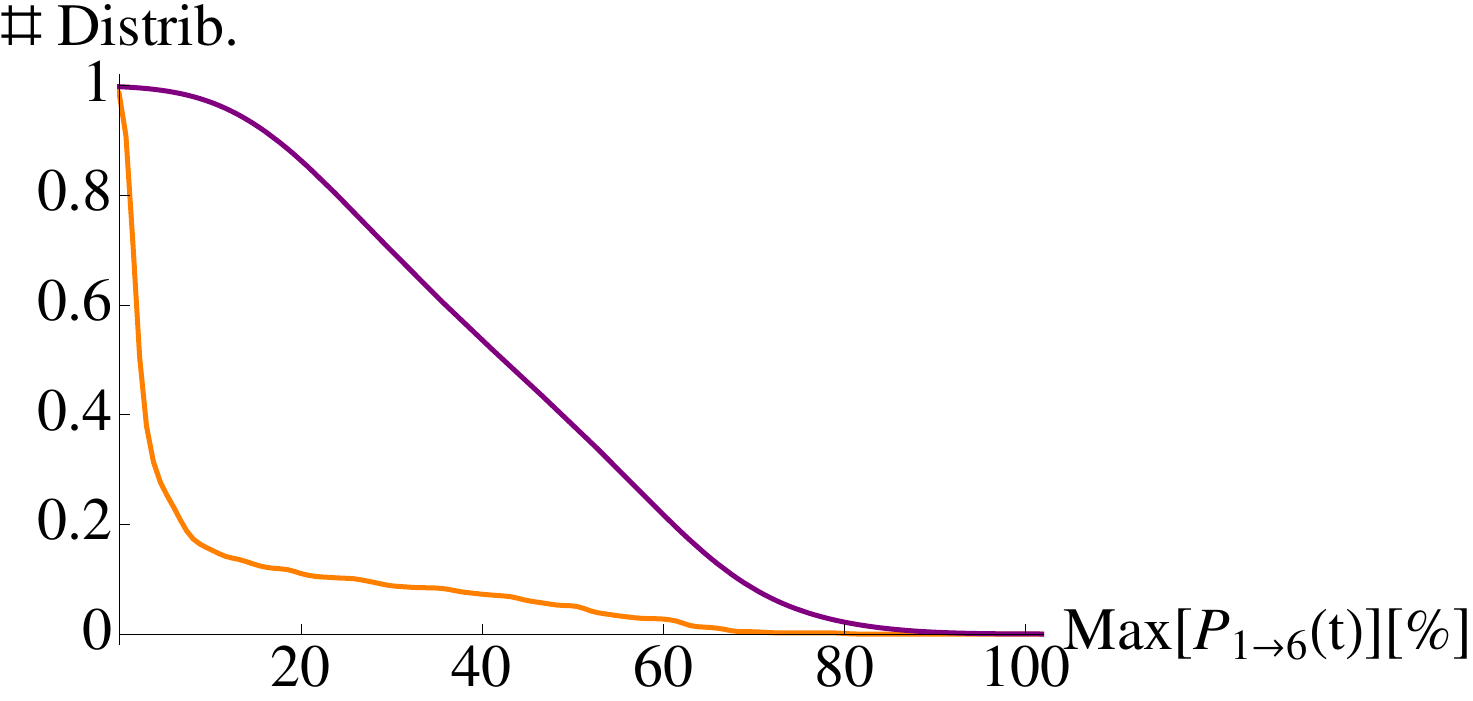} }}
\caption{Cumulative distributions of the maximum probability of transfer for the FMO complex coupled to the 462 different spin distributions with a total of $n$=10 spins at 300K. In the figure on the left(right), the transfer is from site 1 to site 3(6), where the upper curve represents distributions where the initial and final site energies are shifted towards equality, i.e. $n_1=2, n_{3}=0$ ($n_6=4, n_{3}=0$), while the lower curve is for the total distribution set.} 
\end{figure}
%
%
It has been noted that by neglecting the weakest couplings in the FMO Hamiltonian, transport in an individual monomer of FMO can be mapped to a one-dimensional path between chromophores \cite{hoyer10}. We find in our simulations that features of both a uniformly coupled network and a chain contribute to the optimal transfer probabilities for the Hamiltonian $H^{\mathrm{FMO}}_{\mathrm{ex}}$. This can be seen by examining the optimal spin distibutions for transfer through the complex.\\
\\
We have found by studying the properties of fully connected networks that transfer is achieved with high probability when only the initial and final sites have comparable energies, i.e. when $\varepsilon_{I}=\varepsilon_{F}\neq\varepsilon_{j\neq I,F}$. For $H^{\mathrm{FMO}}_{\mathrm{ex}}$, we find that the distributions that achieve this; namely $n_1=2, n_{3}=0$($n_6=4, n_{3}=0$) for transfer from site 1(6), do well for all positions of the remaining spins relative to the total set of distributions, see figure 4.\\
\\
At the same time, having the majority of the environmental spins positioned so as to energetically block further transfer beyond the final site in the effective chain also contributes to optimality. For example, for transfer from site 1 to site 3, the spin distributions and corresponding transfer probabilities that have the three highest transfer probabilities are the following:
\bea
n_1&=&2, n_4=8\ (90\%),\\
n_1&=&2, n_4=6, n_6=2\ (85\%),\\
n_1&=&2, n_4=6, n_7=2\ (85\%),
\eea
where the maximum energy shift associated with 2 spins at the optimal coupling strength of $\gamma_{\mathrm{opt}}\approx 170\mbox{ps}^{-1}$ is then approximately 900cm$^{-1}$.\\
\\
In our model, the value of the bath energy parameter $\alpha$ as compared with $k_{B}T$ determines the influence of thermal fluctuations on the dynamics; when $\alpha/k_{B}T$ is small, the effect of the decoherent interaction with the environmental spins is reduced, while when $\alpha/k_{B}T$ is big, the effect is maximised (see figure 5).\\
\\
\begin{figure}
\centering
\includegraphics[scale=0.7]{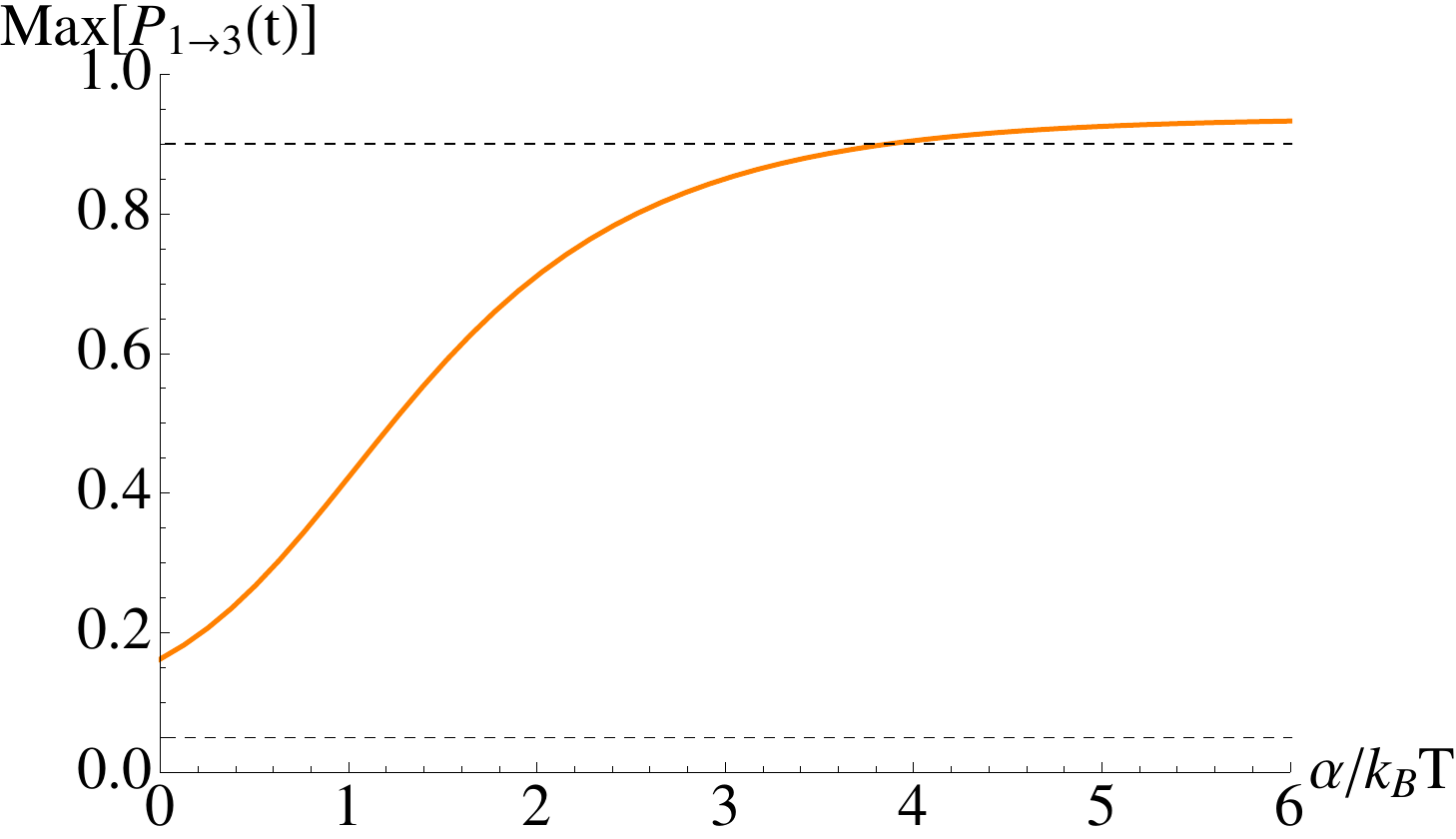}
\caption{Maximum transfer probability from site 1 to 3 in the FMO complex during the first picosecond as a function of $\alpha/k_{B}T$ with $T$=300K. The coupling constant and spin confiuration are chosen to be optimal; $\gamma_{\mathrm{opt}}=170$ps$^{-1}$ and $n_1=2,\ n_4=8$. When $\alpha/k_{B}T$ is small, the maximum transfer probability is not dramatically increased from the case with no bath (the lower dashed line at 5$\%$), while the value of 90$\%$ is attained when $\alpha/k_{B}T$ is increased to 3.82 (the upper dashed line), and beyond that, the optimal value of 94$\%$ is reached.}
\end{figure}
For transfer between sites 1 and 3 of the FMO complex, with sites coupled to the optimal spin distribution, $n_1=2$, $n_4=8$, we find the optimal values of $\gamma$ and $\alpha$ common to both $T=77K$ and $T=300K$ to be $\gamma_{\mathrm{opt}}=170\mbox{ps}^{-1}$ and $\alpha_{\mathrm{opt}}=460\mbox{ps}^{-1}$. At these values, we find that the coherence between sites 1 and 3, defined as
\be
c_{1,3}(t)=\frac{|\rho_{1,3}(t)|^2}{\rho_{1,1}(t)\rho_{3,3}(t)}
\ee
where $\rho(t)$ is the reduced density matrix of the FMO system, is unity for all times and for both temperatures, while ultrafast two-dimensional spectroscopy performed on the FMO complex reveals that the rate of decoherence has a strong dependence on temperature \cite{panit10}. We have therefore set $\alpha=150$ps$^{-1}<\alpha_{opt}$ throughout this work, such that the rates of decoherence show a strong temperature dependence, but at the same time the extent of the effect can be observed.

\section{Conclusion}

The recent detection of quantum coherence in biological systems that are remarkably efficient in transferring excitation energy at physiological temperatures, has led to the proposal of a number of environment-assisted quantum transport models. Here, we have investigated the influence of environmental spins on quantum coherent transfer through a network. We have shown through the derivation of analytic expressions that the transfer probabilities through a fully connected quantum network are improved as a result of decoherent interaction with environmental spins, and that in some cases certain transfer can be achieved. Moreover, this effect is shown to persist at physiological temperatures. We apply this model to the FMO complex, and find that coupling the network sites with environmental spins at physiological temperature improves transport through the network for a vast majority of considered cases. Our results show that features associated with uniformly coupled networks as well as chain-like characteristics of the FMO complex contribute to the optimal transfer efficiencies. These promising results motivate further study of biological transport systems where a spin bath could provide a sufficiently realistic model of the environment and play a fundamental role in the dynamics.

\section*{Acknowledgments}
This work is based upon research supported by the South African Research Chair Initiative of the Department of Science and Technology and National Research Foundation, as well as the National Research Foundation and the Ministry of Education, Singapore. 

\appendix
\section*{Appendix}

For the Hamiltonian $H_{k}$ where $N-k$ of the fully connected network sites, sites $|k+1\rangle,...,|N\rangle$, are coupled to spin baths, and with the initial and final sites chosen from any of the sites $|1\rangle,...,|k\rangle$, the transfer probability is given by 
\be
P_{I\rightarrow F}(t)=\frac{1}{k^2}|1-\langle\phi|\mbox{e}^{-itH_{k}}|\phi\rangle|^2.
\ee
The maximum of the above expression is $4/k^2$; attained when the state $|\phi\rangle$ evolves over time $t$ to the state with the opposite phase, i.e. $-|\phi\rangle$. Motivated by the existence of recurrence properties, we might expect that this maximum can always be achieved. Here, we construct an example of a system which never achieves it, and outline the conditions under which the bound can be approximately reached in the general case.\\
\\
The quantum recurrence theorem states that the state vector for any quantum system with discrete energy eigenstates evolves arbitrarily closely to the initial state, infinitely often \cite{bocchieri57}. On the other hand, there exist quantum systems where the initial state never evolves into the same state but with opposite phase, as required here. For example, the initial state $|I\rangle=(1,0,0)^T$ will never evolve under the Hamiltonian
\be
H=
\left( {\begin{array}{lll}
0&J_{1}&0\\
 J_{1}&0&J_{2}\\
 0&J_{2}&0
 \end{array} } \right)
\ee
into the state $-|I\rangle$, since 
\be
\langle I|\mbox{e}^{-itH}|I\rangle=\frac{J_{2}^2+J_{1}^2\cos{\sqrt{J_{1}^2+J_{2}^2}t}}{J_{1}^2+J_{2}^2}\neq -1, 
\ee
as long as $J_{2}\neq 0$.\\
\\
This specific example can be used to construct instances that coincide with our model. Consider a fully connected network $H_{FCN}$ with intersite coupling $J$, and where equal site energies $\varepsilon_{j}$ are grouped into 3 blocks of size $k_j$:
\be
H_{FCN}=\sum_{j=1}^{k_1}\varepsilon_{1}|j\rangle\langle j|+\sum_{j=k_1+1}^{k_1+k_2}\varepsilon_{2}|j\rangle\langle j|+\sum_{j=k_1+k_2+1}^{k_1+k_2+k_3}\varepsilon_{3}|j\rangle\langle j|.
\ee
Using a unitary operation of the form defined in section 3.2, the Hamiltonian $H_{FCN}$ can be transformed into an effective Hamiltonian of dimension 3, $\tilde{H}_{FCN}$, which when compared with Hamiltonian $H'$ via a unitary that satisfies $V|I\rangle=|I\rangle$, yields conditions on the constants $k_{j}$ and $\varepsilon_{j}$ such that $\langle I|\mbox{e}^{-it\tilde{H}_{FCN}}I\rangle\neq -1$ for all times $t$.\\ 
\\
One such choice of variables is the following: $k_{1}=21$ with $\varepsilon_{1}=0$, $k_{2}=6$ with $\varepsilon_{2}=63J/5$, and $k_{3}=1$ with $\varepsilon_{3}=112J/5$. Scaling out the factor $J$ from the time for convenience, the transfer probability $P_{I\rightarrow F}(t)$ between a pair of sites in the first block is then given by the expression
\bea
P_{I\rightarrow F}(t)&=&\frac{1}{642978}[2091-1350\cos(\frac{42t}{5})+200\cos(\frac{63t}{5})\nonumber\\
&&-216\cos(21t)+625\cos(\frac{126t}{5})-1350\cos(\frac{168t}{5})].
\label{period}
\eea
In figure 5, $k_{1}^2/4$ times the transfer probability, $441/4P_{I\rightarrow F}(t)$, is plotted as a function of time. It can be seen here, and in (\ref{period}), that the function is periodic, and also that the value of 1 is never reached.
\begin{figure}
\centering
\includegraphics[scale=0.6]{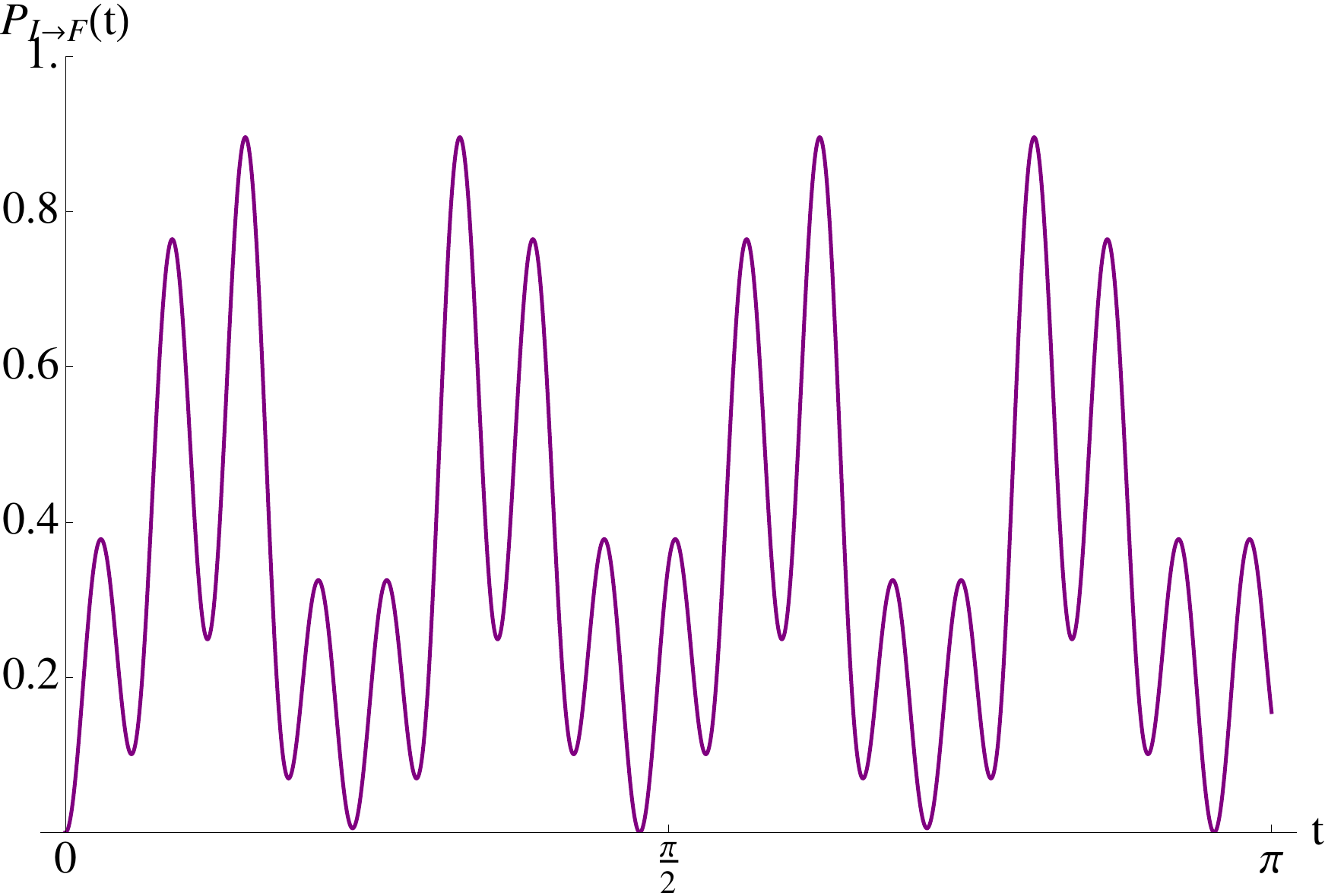}
\caption{Graph of $k_{1}^2/4$ times the transfer probability, $441/4P_{I\rightarrow F}(t)$, for Hamiltonian $H_{FCN}$ with the initial and final sites both from block 1 and with parameters $k_{1}=21$, $\varepsilon_{1}=0$, $k_{2}=6$, $\varepsilon_{2}=63J/5$, and $k_{3}=1$ with $\varepsilon_{3}=112J/5$. It can be seen that the probability never reaches 1, and therefore that the bound of $4/k_{1}^2$ is not achieved for this Hamiltonian $H_{FCN}$.}   
\end{figure}
This example proves that not every network can exactly achieve the optimal transfer, or even get arbitrarily close. Nevertheless, we now discuss how the value $\mbox{Max}[P_{I\rightarrow F}(t)]=4/k^2$ is approximately achieved for almost all general Hamiltonians $H_{k}$.\\
\\
The Hamiltonian $H_k$ given in (\ref{Hk}) for a general homogeneous fully connected network can be partitioned into $m$ blocks indexed by $i$, each of dimension $k_{i}$ and with site energies $\varepsilon_{i}=n_{i}B_i$, where $B_i=-\gamma_i/2$. Transforming the Hamiltonian $H_{k}$ into the basis of uniform superposition across each block (via the unitary of section 3.2), and then moving to the rotating basis, i.e. transforming each amplitude $a_{i}$ as 
\be
a_{i}\rightarrow a_{i}\mbox{e}^{-it(k_{i}J+n_{i}B_i)},
\ee
yields an effective Hamiltonian $H_{k}^{eff}$ of dimension $m$, with diagonal elements all 0, and off-diagonal elements of the form 
\be
J\sqrt{k_{n}k_{m}}\mbox{e}^{-it((k_{n}-k_{m})J+(n_{n}B_n-n_{m}B_m)}. 
\ee
In the rotating wave approximation, over long times each matrix element undergoes many oscillations and can therefore be approximated to 0, with the best approximation here achieved at times for which 
\be
\int_{0}^{t}\mbox{e}^{-it[(k_n-k_m)J+(n_{n}B_n-n_{m}B_m)]}dt=0\ \mbox{for all}\ n,m.
\ee
Hence, the amplitude remains (approximately) in the first block, but with an evolving phase that at some time becomes -1.\\
\\
This proof only fails in exceptional cases where some of the couplings of the effective Hamiltonians do not have a time dependent term, and cannot be averaged away. For example, when the initial and final sites are in block 1, and 
\be
(k_{1}-k_{m})J+(n_{1}B_1-n_{m}B_m)= 0, 
\ee
with $n_{1}B_1\neq n_{m}B_m$, it could be that the phase never evolves to -1. However it is sufficient for $J/B_i$ to be irrational to avoid these cases, which means the exceptions are of measure 0, and the upper bound can be achieved in almost all cases.  
\setcounter{section}{1}

\section*{References}

\end{document}